\documentclass{PoS}

\usepackage{amsmath}
\usepackage{graphicx}
\usepackage{mathrsfs}

\title{On the infrared behaviour of QCD Green functions in the Maximally 
Abelian gauge}

\ShortTitle{On the infrared behaviour of QCD Green functions in the 
Maximally Abelian gauge}

\author{\speaker{Reinhard Alkofer}%
       \\
       Institut f\"ur Physik, 
       Karl-Franzens-Universit\"at,
       Universit\"atsplatz 5,
       A-8010 Graz, Austria\\
       E-mail: \email{reinhard.alkofer@uni-graz.at}}
\author{Markus Q.\ Huber\\
        Institut f\"ur Kernphysik,
        Technische Universit\"at Darmstadt,
        Schlossgartenstr. 9,
        64289 Damstadt, Germany\\
	E-mail: \email{markus.huber@physik.tu-darmstadt.de}}
\author{Valentin Mader\\
        Institut f\"ur Physik, 
        Karl-Franzens-Universit\"at,
        Universit\"atsplatz 5,
        A-8010 Graz, Austria\\
        E-mail: \email{valentin.mader@uni-graz.at}}
\author{Andreas Windisch\\
        Institut f\"ur Physik, 
        Karl-Franzens-Universit\"at,
        Universit\"atsplatz 5,
        A-8010 Graz, Austria\\
        E-mail: \email{andreas.windisch@uni-graz.at}}

\abstract{
Functional equations like exact renormalisation group and  Dyson-Schwinger
equations have contributed to a better understanding of non-perturbative
phenomena in quantum field theories in terms of the underlying Green functions.
In Yang-Mills theory especially the Landau gauge has been used, as it is the
most accessible gauge for these methods. In the maximally Abelian gauge first
results have been obtained which are very encouraging because Abelian infrared
dominance has been found: The Abelian part of the gauge field propagator is
enhanced at low momenta and thereby dominates the dynamics in the infrared. 
Also the ambiguity of two different types of solutions (decoupling and scaling)
exists in both gauges. It is demonstrated how the two solutions are related in
the maximally Abelian gauge. As in all two-point Dyson-Schwinger equations of
the MAG the  infrared dominant diagrams are sunset diagrams, in addition, a
BPHZ  regularisation and renormalisation of a test system with a sunset-like 
diagram is presented.}

\FullConference{International Workshop on QCD Green's Functions, 
                Confinement and Phenomenology\\
		 5-9 September 2011\\
		 Trento, Italy}

\begin{document}

\section{Motivation: The dual superconductor picture of confinement}

A generally agreed upon explanation of confinement is still lacking. As a matter
of fact, the last confinement conference has hosted a panel discussion with the
title ``What {\it don't} we know about confinement?'' \cite{Alkofer:2010ue}.
Actually it turns out that there are many unanswered questions, and
we do not even know if we already know all questions. 
Within the many suggested different scenarios the dual 
superconductor picture \cite{Mandelstam:1974pi,'tHooft:1975pu} plays a special
role. It has been with us for now almost forty years due to its appealing
physical nature of an explanation of confinement. Nevertheless, it proved to be
very hard to substantiate this scenario. 

The dual superconductor picture exploits an analogy to the Mei\ss ner-Ochsenfeld
effect  in type-II superconductors: The vacuum is assumed to contain condensed
chromomagnetic monopoles\footnote{Chromomagnetic monopoles are gauge independent
objects \cite{Bonati:2010tz}, nevertheless,  their detection in lattice
configurations in general  depends on the choice of gauge
\cite{Bonati:2010bb}.}  which squeeze the chromoelectric field lines between two
colour charges  into a flux tube. These vortex like structures can be identified
best after  an appropriate choice of gauge which singles out the Cartan
subalgebra of the gauge group's Lie algebra. Several such partial gauge fixings
have been introduced \cite{'tHooft:1981ht}.  The most widely used one is the
maximally Abelian gauge (MAG):  In this gauge the components of the gauge field
outside the Cartan subalgebra are minimized.  As the corresponding generators
are off-diagonal matrices, the corresponding gluon field components are called
{\em off-diagonal gluon fields} in contrast to the {\em diagonal} ones,
which correspond to the gluon field in the Cartan  subalgebra.

If the dual superconductor picture is correct the diagonal gluons must
dominate the infrared (IR) properties of a Yang-Mills theory in the confinement
phase. This follows directly from the fact that also the chromomagnetic
monopoles live in the Cartan subalgebra \cite{Ezawa:1982bf}. A detailed
understanding of this so-called hypothesis of Abelian IR dominance is missing. 
In terms of MAG Yang-Mills Green functions one possible realization is an IR
finite behaviour for all three propagators of the theory  (diagonal and
off-diagonal gluons as well as ghosts) such that the  largest value is assumed
for the diagonal gluon propagator.  Attributing to the inverse of the propagator
at vanishing momentum a screening mass, $m^2\propto D^{-1}(0)$, it is evident
that in the deep IR the off-diagonal gluon fields and the ghost decouple from
the IR dynamics which is then dominated by the diagonal gluons.  Such a behaviour
has been found in lattice Monte-Carlo studies (see, {\it e.g.},
refs.~\cite{Bornyakov:2003ee,Mendes:2008ux}),  in the refined
Gribov-Zwanziger framework \cite{Capri:2008ak,Capri:2010an}, and 
in a so-called replica model of the MAG \cite{Capri:2011ki}.
Another possible realization of Abelian IR dominance would
be a diverging diagonal gluon propagator. Such a behaviour turns out to be a
potential solution of functional equations by employing an IR
scaling analysis of the MAG \cite{Huber:2009wh,Huber:2010ne}. 
This is reminiscent of  the coexistence of a scaling solution with an
one-parameter family of decoupling solutions  for functional equations in other
gauges. Such a multitude of solutions 
is also known from the Coulomb \cite{Szczepaniak:2001rg,Epple:2007ut} and the
Landau gauges \cite{Boucaud:2008ji,Fischer:2008uz}.

To fix the notation we will briefly summarize the concept of the MAG in the
following section. We will then discuss  the scaling analysis as well as the
coexistence of solutions.  As these results point towards a dominance of
two-loop terms in the propagator Dyson-Schwinger equations, we will report on our
efforts to regularize sunset and squint diagrams such that they can be
efficaciously included in a self-consistent solution of Dyson-Schwinger
equations.

\section{Fundamentals of the maximally Abelian gauge}
\label{sec:MAG}

To define the MAG of an SU(N) Yang-Mills theory one needs to split the gauge 
field into diagonal and off-diagonal components:
\begin{align}
 A_\mu=T^i A^i_\mu+T^a B^a_\mu,
\end{align}
where the $T^i$ are the $N-1$ generators of the Cartan subalgebra of  $SU(N)$
with $[T^i,T^j]=0$.  In the physically interesting case of $SU(3)$  the diagonal
generators are $T^3$ and $T^8$. It is a widely employed notation  to use the
indices $i,j,\ldots$ for diagonal generators only, and  $a,b,\ldots$ for
off-diagonal ones. The indices $r,s,\ldots$ are used if both types, diagonal and
off-diagonal, generators are present within one relation. Furthermore, we use
$A$ for the diagonal  and $B$ for the off-diagonal gluon fields. 

The herewith introduced splitting of gluon fields has direct consequences for 
the  possible interactions between the now different components of the 
gluon fields. 
Starting from the standard commutation relation of the generators,
\begin{align}\label{eq:structure-constants}
 [T^r,T^s]=i\,f^{rst} T^t,
\end{align}
one can directly see that only three off-diagonal or two off-diagonal and one
diagonal fields can interact.  For the gauge group $SU(2)$ there are more
restrictions because  there exist only two off-diagonal generators and therefore
the first possibility cannot be realized. The number of interaction vertices in
$SU(N>2)$ is then larger than in $SU(2)$.  In summary the pure Yang-Mills part
has the following types of interaction vertices: 
$ABB$, $AABB$ and $BBBB$ for $SU(2)$ and additionally  $BBB$ and $ABBB$ for
$SU(N>2)$.

The MAG is tuned to minimize the norm of the off-diagonal gluon fields
in order to make the effect of the diagonal part most pronounced. 
The extrema of the functional 
\begin{align}
\frac1{2} \int dx \, B^a_\mu(x) B^a_\mu(x) 
\end{align}
are taken if $D^{ab}_\mu B_\mu^b=0$. 
Hereby the covariant derivative contains only the diagonal gluon field:
\begin{align}
D_\mu^{ab}:= \delta^{ab}\partial_\mu+g\,f^{abi} A_\mu^i.
\end{align}
To employ functional methods 
the remaining $U(1)^{N-1}$ symmetry of the action is conveniently 
fixed to the Landau gauge,
$\partial_\mu A_\mu^i=0$. Since the field $A^i$ is Abelian the corresponding
Faddeev-Popov ghosts decouple and only the ghosts of the non-diagonal sector 
need to be taken into account.
Hereby, a quartic ghost self-interaction term in the action is required to
maintain  renormalisability \cite{Min:1985bx,Capri:2005zj}. The interactions stemming from the gauge fixing and the
renormalisability requirement are therefore 
$Acc$, $AAcc$, $BBcc$, $cccc$, $Bcc$, and 
$ABcc$. The two latter interaction vertices vanish for $SU(2)$.
This finally leads to the complete action of Yang-Mills theory in MAG:
\begin{align}
S_{MAG}&= \int dx \Big(\frac{1}{4}F_{\mu\nu}^i F_{\mu\nu}^i+
\frac{1}{4}F_{\mu\nu}^a F_{\mu\nu}^a +
\bar{c}^a D_\mu^{ab}D_\mu^{bc} c^c-
g\,f^{bcd}\bar{c}^a D_\mu^{ab} B_\mu^c c^d 
-g^2\,\zeta f^{abi}f^{cdi} B_\mu^b B_\mu^c \bar{c}^a c^d+\nonumber\\
&\qquad +\frac1{2\alpha}(D_\mu^{ab} B_\mu^b)^2
+\frac{\alpha}{8} g^2 f^{abc}f^{ade} \bar{c}^b c^c \bar{c}^d c^e
-\frac{1}{2} g\,f^{abc} (D_\mu^{ad} B_\mu^d) \bar{c}^b c^c + \nonumber \\
&\qquad +\frac1{4}g^2\alpha f^{abi}f^{cdi}\bar{c}^a\bar{c}^b c^c c^d+
\alpha\frac1{8}g^2 f^{abc}f^{ade} \bar{c}^b \bar{c}^c c^d c^e+
\frac1{2\xi}(\partial_\mu A_\mu^i)^2\Big),
\end{align} 
where $\alpha$ is the gauge fixing parameter of the off-diagonal  and 
$\xi$ the one of the diagonal sector. 

For an action containing three fields with eleven interaction vertices   the
derivation of its Dyson-Schwinger equations (DSEs) or functional renormalisation
group equations (FRGEs) is very lengthy. To this end we employ the computer
algebra package \textit{DoFun} \cite{Huber:2011qr}, the successor of
\textit{DoDSE} \cite{Alkofer:2008nt}\footnote{ Recently, the program 
\textit{CrasyDSE } has become available \cite{Huber:2011xc}. 
It provides a framework for solving DSEs numerically and can directly be 
combined with \textit{DoFun}.}.  
Two new features of  \textit{DoFun} as compared to \textit{DoDSE} are the 
inclusion of FRGEs and the derivation of the Feynman rules from a given action. 
In addition,  the complete algebraic form of the integrands can now be obtained
so that computations directly with \textit{Mathematica} are possible. This was
very helpful for the calculations of the IR leading diagrams of the MAG
\cite{Huber:2010ne} and the Gribov-Zwanziger action \cite{Huber:2009tx}. For the
latter the application of \textit{DoFun} helped to identify a unique solution
\cite{Huber:2010cq}: Only with the help of a computer algebra system one is able
to show that one of the two possible scaling solutions  found in
\cite{Huber:2009tx} does not yield a numerical solution for the infrared
exponents and can thus not be realized.

\section{Infrared analysis}
\label{sec:scalingAnalysis}

Functional methods provide equations for the fully dressed Green functions.
Therefore these equations are applicable to the IR regime of Yang-Mills 
theories and provide, at least in principle, information about the IR behaviour
of the theory  \cite{Alkofer:2000wg,Fischer:2006ub}. 
The gauge investigated best is the Landau gauge. Here two 
types of solutions have emerged from the analysis of functional equations: 
The scaling solution
\cite{Fischer:2008uz,Huber:2009tx,Alkofer:2000wg,Fischer:2006ub,vonSmekal:1997vx,Watson:2001yv,Pawlowski:2003hq,Lerche:2002ep,Alkofer:2008jy,Alkofer:2011pe,Alkofer:2004it}
and the one-parameter family of decoupling solutions 
\cite{Boucaud:2008ji,Fischer:2008uz,Alkofer:2008jy,Dudal:2007cw,Aguilar:2008xm}.
The former is characterized by an IR vanishing gluon propagator, an IR
enhanced ghost propagator and IR enhanced three- and four-gluon
vertex functions, whereas the latter possesses only IR finite 
dressing functions except for the gluon dressing \cite{Alkofer:2008jy}. 
Both solutions are connected by the choice
of a  renormalisation condition which is a needed input to solve the equations
\cite{Boucaud:2008ji,Fischer:2008uz}. 
The difference only affects the deep IR behaviour of the
propagators and vertex functions, while all solutions agree for intermediate
momenta well below the scale where perturbation becomes valid\footnote{As
perturbation theory is included in the functional equations all solutions,  of
course, agree on the multi-GeV scale.}. Up to now no
calculation exists where these two types of solutions generate a difference 
for physical quantities: 
In ref. \cite{Braun:2007bx} a confining Polyakov loop potential was derived 
for both types, the
confinement and the chiral transition temperatures were found to
coincide for scaling and decoupling \cite{Fischer:2009gk}, 
and there are hints that even meson masses are
independent of this choice \cite{Blank:2010pa}. This substantiates the
interpretation of this additional type of boundary condition as a 
non-perturbative gauge fixing
parameter as in the Landau-$B$ gauges of ref. \cite{Maas:2009se}. 
It is worthwhile to mention that this dichotomy also exists in Coulomb gauge
\cite{Szczepaniak:2001rg,Epple:2007ut}.

One major result to be presented here is that  in the MAG one also obtains this
two qualitatively different types of  solutions, and that there exists a similar
connection between them. For the IR analysis we make the ansatz that all
dressing functions  follow a power law in the deep IR, {\it e.g.}, for a
propagator one uses the ansatz $D(p)=c(p^2)/p^2$ and
$c^{IR}=d\cdot(p^2)^{\delta}$,  where $\delta$ is the related infrared exponent
(IRE). As we are in a first step interested in the  qualitative behaviour we
want to determine the  different IREs. Therefore we exploit the very welcome  
possibility to shift the analysis to the level of the IREs only. This type of
IR analysis uses the fact that all integrals are dominated by low momenta if
the external momenta are low. Thus, for the purpose of obtaining 
equations for the IREs one can replace all propagators and vertices
by the corresponding IR expressions \cite{Alkofer:2004it,Huber:2007kc}. Then
one counts all exponents of momenta and calculates by this procedure 
the IRE of any given diagram.

The most elaborate method to derive the scaling relations of the IREs is the
combined use of Dyson-Schwinger and Functional Renormalisation Group equations.
This was first done for the Landau gauge in 
Refs.~\cite{Fischer:2006vf,Fischer:2009tn}. 
This method can be generalized rather generically so that one can reduce the 
infinitely many
equations for the IREs  to a rather small number of relations 
between the IREs \cite{Huber:2009wh,Huber:2010ne}.

Without loss of generality, for the MAG at $\xi=0$ the propagators can be 
parametrized as
\begin{align}
 D_A^{ij}(p^2)&=\delta^{ij}\frac{c_A(p^2)}{p^2}
 \left(g_{\mu\nu}-\frac{p_\mu p_\nu}{p^2}\right),\\
 D_B^{ab}(p^2)&=\delta^{ab}\frac{c_B(p^2)}{p^2}
 \left(g_{\mu\nu}-(1-\alpha)\frac{p_\mu p_\nu}{p^2}\right),\\
 D_c^{ab}(p^2)&=-\delta^{ab}\frac{c_c(p^2)}{p^2} .
\end{align}
The following notation for the power laws of the dressing functions 
in the IR is used:
\begin{align}
 c_A(p^2)&\overset{p^2\rightarrow 0}{=} d_A \cdot (p^2)^{\delta_A},&\quad
 c_B(p^2)&\overset{p^2\rightarrow 0}{=} d_B \cdot (p^2)^{\delta_B},&\quad
 c_c(p^2)&\overset{p^2\rightarrow 0}{=} d_c \cdot (p^2)^{\delta_c}.
\end{align}
As it is not possible to set the gauge fixing parameter of the off-diagonal 
part directly to zero the longitudinal part of the off-diagonal propagator 
could in principle acquire an own dressing function. 
However, a careful analysis reveals that no new IRE is generated.
Based on this we assume in the following infrared analysis only one common
dressing function for both tensors.

The main result is that there exists  only one consistent scaling relation.
It reads \cite{Huber:2009wh}: 

\begin{align}\label{eq:MAGScalingRelation}
 \kappa_{MAG}:=-\delta_A=\delta_B=\delta_c\geq0.
\end{align}
This result for the IREs especially implies that  {\bf the diagonal gluons are
IR enhanced} and the off-diagonal ones are IR suppressed. An upper bound for
the parameter $\kappa_{MAG}$ can be obtained by demanding  well-defined
Fourier transformations of the propagators \cite{Lerche:2002ep}:
$\kappa_{MAG}<1$. The IREs for the vertex functions are such  that these
functions  become more IR divergent when the number of off-diagonal legs
increases \cite{Huber:2009wh}. This applies to the MAG in $SU(2)$ and
$SU(N>2)$. It is an interesting result in itself  that no differences for the
IREs have been found. The additional interactions for larger gauge groups do
neither spoil the  $SU(2)$ relations  nor do they allow an additional solution.

As understood by now the treatment of the bare two-point functions play a
decisive role in the respective Dyson-Schwinger equations.  To allow for a
scaling solution at least one of the zero-momentum values of these Born terms
has to be canceled by quantum effects. This is well known for the scaling
solution in Landau gauge: The bare ghost two-point function  vanishes due to
the choice of a corresponding renormalisation condition
\cite{vonSmekal:1997vx,Fischer:2008uz,Zwanziger:2001kw}.  To put no prejudice
on the system we allow all Green functions to become IR divergent. However, it
turns out that {\bf the only consistent solution is the one with an IR enhanced
diagonal gluon propagator}. As a first remark, we want to emphasize that we
count this as strong evidence that also in the decoupling solution the diagonal
gluon propagator is the IR dominant quantity. Second, we point out that the
value of the diagonal two-point function at zero momentum can serve as an
additional gauge fixing parameter as the ghost two-point function does in the
Landau-$B$ gauges of Ref.~\cite{Maas:2009se}. To be precise, an IR
finite diagonal gluon propagator implies IR finite off-diagonal propagators.
The smaller the value in the renormalisation condition becomes the closer one
is to the scaling solution. 

In the MAG, the reason for this entanglement is the quartic interactions
between diagonal and off-diagonal fields. If the diagonal gluon propagator is
IR finite, the corresponding tadpole diagrams are proportional to the inverse
of the IR value of the diagonal gluon propagator. Via this mechanism IR finiteness
of all  two-point dressing functions is implied.  
Tuning the zero momentum value of the
diagonal gluon propagator to zero by an appropriate renormalisation condition
switches  then to the scaling  solution. Note that this is a different
situation than in Landau (or Coulomb) gauge: There is no quartic interaction
between the ghost and the gluon in the Landau gauge, and consequently an IR
finite gluon propagator does not enforce an IR finite ghost propagator.

This IR analysis  is, of course, not compelling evidence that the scaling
solution exists (at least, in the mathematical sense). A  numerical
calculation is, however, enormously more complicated  than in Landau gauge.
This is implied by the nature of the MAG scaling solution: Two-loop
diagrams are the IR leading ones.  Therefore, for a consistent numerical
treatment a more elaborated truncation scheme has to be developed.  Continuity
arguments then imply that  even for the MAG decoupling solution  a
straightforward one-loop truncation likely will miss the essential terms.

\section{Determination of infrared exponents}

\begin{figure}[tb]
\begin{center}
\includegraphics[width=0.45\textwidth]{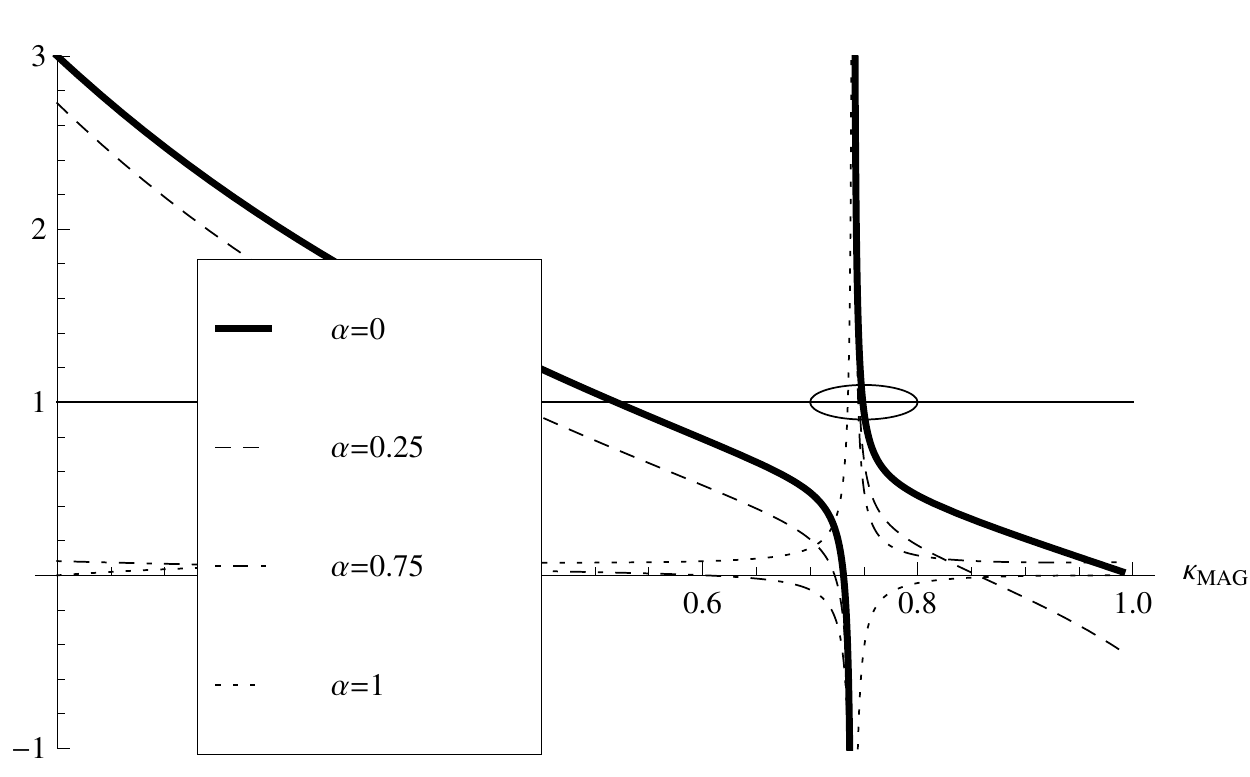}
\includegraphics[width=0.45\textwidth]{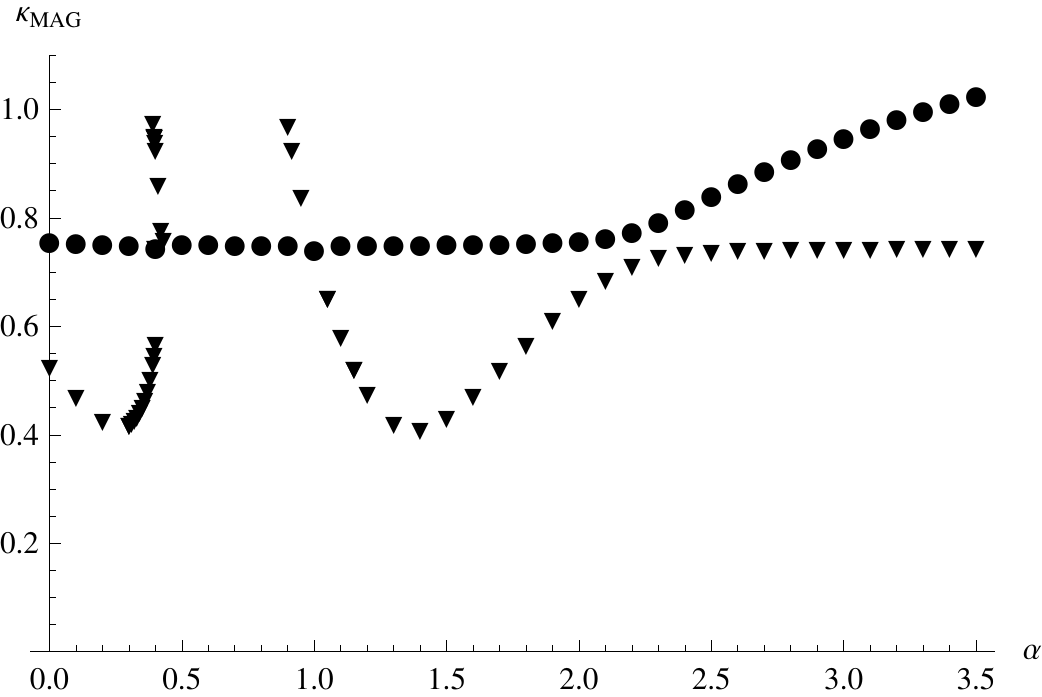} 
\caption{
\label{fig:kappaMAG} 
In the left panel the r.h.s.~of eq.\ (4.4) 
is  shown for
several values of the gauge fixing  parameter  $\alpha$. Crossings with the
horizontal line represent solutions for  $\kappa_{MAG}$. In the right panel the
two solution branches are displayed as function of $\alpha$.}  
\end{center} 
\end{figure}

Before a numerical solution of the MAG functional equations is attempted it is
instructive to calculate the parameter $\kappa_{MAG}$ and therewith the
IREs\footnote{In this calculation the use of the 
program \textit{DoFun}  \cite{Alkofer:2008nt,Huber:2011qr} was vital
since the intermediate expressions appearing there are many
pages long.}. To do
this one can restrict oneself to  the IR leading diagrams. Again  an ambiguity
is found because the role of the squint diagrams cannot be determined by the IR
analysis alone. In the first step we restricted ourselves to the sunset
diagrams, and the accordingly  projected equations reduce in the IR to
\begin{align} 
d_A^{-1}&=-X^A_{AABB}(p^2,\kappa_{MAG}) d_A d_B^2 -
X^A_{AAcc}(p^2,\kappa_{MAG})  d_A d_c^2,\\
d_B^{-1}&=-X^B_{AABB}(p^2,\kappa_{MAG})  d_A^2 d_B,\\
d_c^{-1}&=-X^c_{AAcc}(p^2,\kappa_{MAG})  d_A^2 d_c.  
\end{align}  
The
quantities $X(p^2,\kappa_{MAG})$ denote the sunset integrals  without the
coefficients from the propagator power laws. The superscript gives the
corresponding Dyson-Schwinger equation, and the subscript the bare vertex
contained in the diagram. Using the invariant combinations $ I_1:=d_A^2 d_B^2$
and $I_2:=d_A^2 d_c^2$ the three equations can be combined:  
\begin{align}
\label{eqForKappa}  
1= \frac{X^A_{AABB}(p^2,\kappa_{MAG})}{
X^B_{AABB}(p^2,\kappa_{MAG})} +
\frac{X^A_{AAcc}(p^2,\kappa_{MAG})}{X^c_{AAcc}(p^2,\kappa_{MAG})}.  
\end{align}
With  $X(p^2,\kappa_{MAG})$ being known this equation yields the solution(s)
for $\kappa_{MAG}$. For their computation  the dressed four-point functions are
needed as input. Their power law behaviour is known to be IR constant, but the
corresponding tensor structures are not available.  We employ the reasonable
assumption that the tree-level structures reflect the general properties. (For
these tree-level expressions see, {\it e.g.}, \cite{Huber:2010ne}.)  The last
unspecified quantity is the gauge fixing parameter $\alpha$  of the off-diagonal
part . In the left panel of Fig.\  \ref{fig:kappaMAG} the r.h.s.\ of
eq.~(\ref{eqForKappa}) is displayed for several values of $\alpha$. It turns
out that there are  solutions with $0.7<\kappa_{MAG}<0.8$ for all reasonable
values of $\alpha$. The right panel displays this directly: As a function of  
$\alpha$  there are  two solution branches, 
with one solution always close to $\kappa_{MAG}\approx 0.74$. The independence of 
$\alpha$  is a
surprisingly positive outcome, and it  even persists for rather high $\alpha$. 

We want to emphasize that obtaining a  solution for  $\kappa_{MAG}$ is a highly
non-trivial result: The IR analysis depends only on the combinatorics of 
Feynman diagrams whereas the computation  of $\kappa_{MAG}$ relies on all the
little details such  as Lorentz and colour structure.

\section{Renormalisation of two-loop terms in Dyson-Schwinger equations}

As described above the challenging task for a consistent truncation of the 
Dyson-Schwinger equations of MAG Yang-Mills theory rests on the inclusion of
two-loop terms. In this respect it turns out that the renormalisation of the
latter  (which is already a considerable task in perturbation theory)  provides
the first serious obstacle for a self-consistent solution. As it is well-known
truncations do interfere with renormalisation. This leads to so-called "spurious
divergences" as truncation artefacts.  These spurious UV divergences have been
observed in many practical calculations, and different techniques to identify or
even avoid them have been proposed.  For example, for the ghost-gluon system in
Landau gauge the Brown-Pennington-projector \cite{Brown:1988bn} can be applied
to keep only the renormalisable terms.

In self-consistent calculations on the one-loop level it has turned out that a
quite successful technique consists in  applying a momentum subtraction scheme
including only the finite and logarithmically divergent self-energies. The
renormalisation conditions are imposed onto the dressing functions at the
subtraction scale \cite{Fischer:2002hna,Fischer:2003zc}. Multiplicative
renormalisability is then also one of  the guiding principle for the construction
of ans\"atze for the truncated Green functions \cite{Fischer:2008uz}. These
techniques, however, have only been applied to one-loop truncations \footnote{
Two-loop terms have been addressed in ref.~\cite{Bloch:2003yu}, but in this study
the two-loop terms have been approximated by two one-loop integrals.}. 

From perturbation theory we know that the structure of divergences in two-loop
terms can be much more complicated then in the one-loop case. UV divergences can now
be overlapping and/or nested.  In the past many techniques have been developed
to remove such UV divergences. Among them are dimensional
regularisation and the BPHZ-procedure using Taylor-subtraction
\cite{Bogoliubov:1957gp,Bogolyubov:1980nc,Hepp:1966eg,Zimmermann:1969jj}.
Dimensional regularisation belongs to the most spread regularisation procedures
in perturbative calculations since it preserves gauge invariance intrinsically.
Unfortunately, in self-consistent calculations this approach is
numerically very demanding \cite{Schreiber:1998ht}. 

The strength of the BPHZ procedure is that it is a clear, well-known and tested
method in perturbation theory. It handles overlapping and nested divergences and
can render any Feynman diagram finite. In addition, it is free of analytical
functions which are numerically expensive to calculate. The general idea to
include the BPHZ method into DSE calculations is as follows: Every specific
diagram appearing in a (truncated)  DSE can be made finite using Zimmermann's
forest formula, {\it i.e.}, all UV divergences are removed, and the integral is
then cut-off independent. The renormalisation is then again, as in one-loop
calculations, performed in a MOM-scheme. 

To exemplify the method we consider a generic propagator DSE. All terms depend on
the external momentum $p$, on a renormalisation scale $\mu$  and a cut-off scale
$\Lambda$. Given some renormalisation constants $Z_i$ one has:
\begin{equation}\label{DSEgen}
 D^{-1}(p;\mu)\,p^2 = Z_3(\mu,\Lambda)\, p^2 + Z_1(\mu,\Lambda)\Pi(p;\Lambda)   \;,
\end{equation}
where $ D(p;\mu)$ denotes the propagator dressing function and $\Pi(p;\Lambda)$
the self-energy term.

Eq.~\eqref{DSEgen} is finite by definition. The two factors $Z_3$ and $Z_1$,
however, are totally unknown. They have to be determined in the renormalisation
process. The dependence of the  self-energy terms on the cut-off is  removed by
the forest formula which introduces a new scale into the calculation, the
subtraction point $s$:   
$ Z_1(\mu,\Lambda)
\Pi(p;\Lambda) \rightarrow Z_1(\mu,s) \tilde \Pi(p;s) \,. $

Imposing multiplicative renormalisability  one obtains conditions for  the
dressing functions of the truncated Green functions. One can then absorb the
renormalisation constant $Z_1$ in the ansatz for the truncated Green function
\cite{Fischer:2002hna,Fischer:2003zc,Fischer:2008uz} and write
$
 Z_1(\mu,s) \tilde\Pi(p;s) \rightarrow \tilde\Pi(p;\mu) \,.
$

Having obtained the properly BPHZ regularized DSE one can perform the
renormalisation procedure in a MOM scheme.  In the last equation the
renormalisation constant $Z_3$ is the only unknown quantity left. To remove it
we subtract at a renormalisation scale $p^2 = \mu^2$  and get
\begin{equation}\label{DSEren}
 D^{-1}(p;\mu) = D^{-1}(\mu;\mu)  + \tilde \Pi(p;\mu) - \tilde \Pi(\mu;\mu) \,.
\end{equation}
The self-energy term can now be calculated and gives a definite cut-off 
independent result. The term $D^{-1}(\mu;\mu)$ is defined by the 
renormalisation condition.


\begin{figure}[t]
\centering
 \includegraphics[scale=0.25]{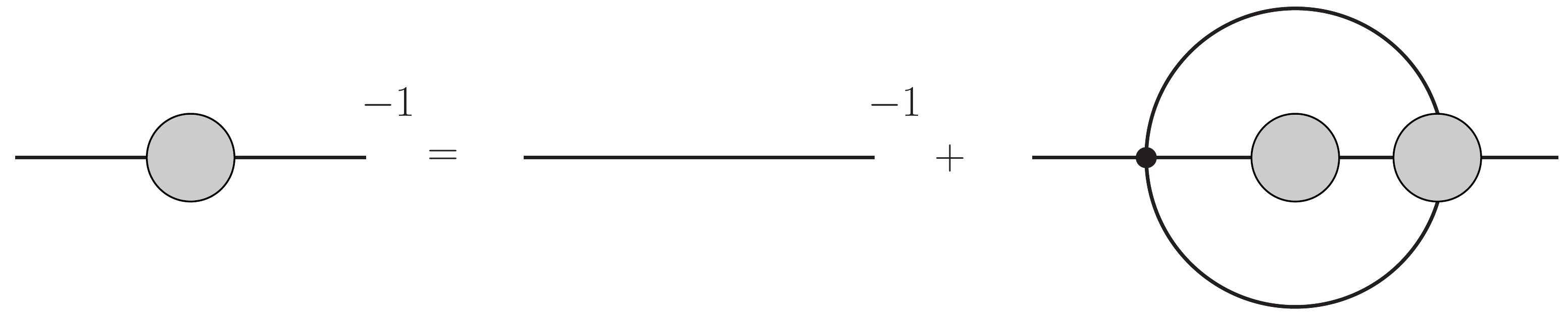}
\caption{A simple self-consistent equation to test the algorithm described in the text.\label{picDSE}}
\end{figure}

To test the procedure we will apply it to the simple DSE-like integral equation
with scalar propagators in four dimensions depicted in Fig.\ref{picDSE}.  Its
divergences are treated  by applying a BPHZ  subtraction to its integrand. The
regularized integrand is obtained by applying the forest formula to the
unregularised integrand: 
$
\tilde I_\Pi=\sum_{\mathscr{U}\in\mathscr{F}_\Pi}
\prod_{\xi\in\mathscr{U}}(-t^{d(\xi)})I_\Pi.
$
The sum in this equation runs over all elements of the family of
forests $\mathscr{F}_\Pi$ of the self-energy diagram $\Pi$,
while $\xi$ addresses the subdiagrams appearing in the subsets $\mathscr{U}$.
The operator $t^{d(\xi)}$ is the Taylor subtraction operator
up to the order of the superficial degree of divergence of the subelement $\xi$,
and is defined by
$
t^{n}=\sum_{i=0}^n\frac{p^n}{n!}
\left[\frac{\partial^n}{\partial p^n}\right]_{p^2=0}.
$
For the sunset-diagram we then obtain a finite and cut-off independent 
integrand given by
$
\tilde I_\Pi=(1-t^{d(\Pi)})I_\Pi.
$

\begin{figure}[t]
\centering
\includegraphics[scale=0.35]{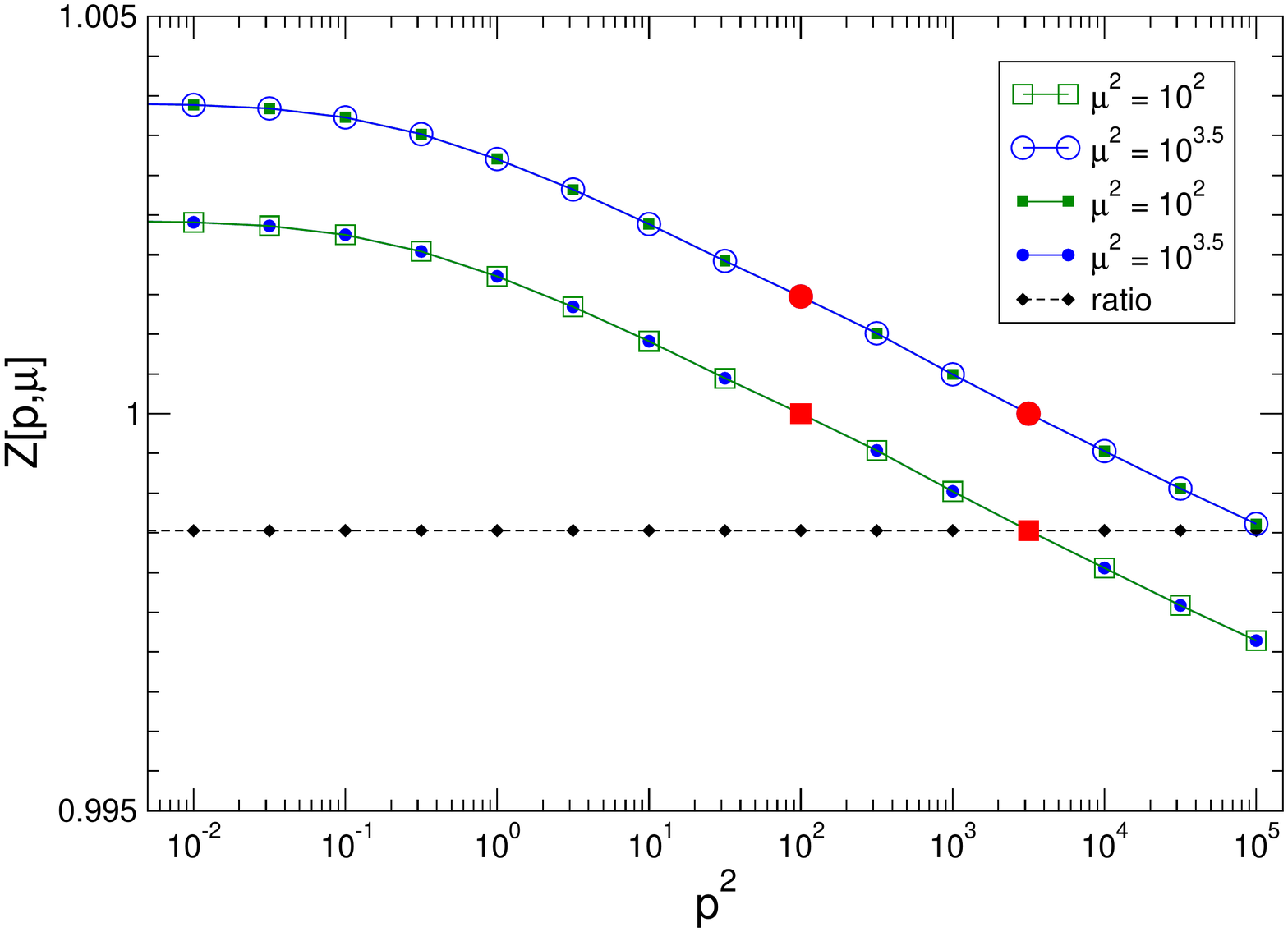}
\caption[]{Solution of the equation depicted in Fig.~\ref{picDSE}. 
Multiplicative renormalisability is evident as the two obtained solutions differ 
only by the constant ratio. The renormalisation points are marked in red.}
\label{solution}
\end{figure}

Next, we have to determine the superficial degree of divergence which is given by
two (quadratically divergent). We thus have to Taylor-subtract up to second order
to get the regularized integrand.  We choose a particular momentum routing
through the three propagators of the diagram from which only the  middle one is
considered dressed. The upper and the lower propagator carry a momentum of
$\frac{1}{2}p+q$ and $\frac{1}{2}p+k$ respectively, where all momenta flow from
the left to the right and where $p$ is the external momentum. The propagator in
the middle carries thus a momentum of $-k-q$. For the 4-scalar vertex we use an
ansatz leading then to the unregularised integral
\begin{eqnarray}
\Pi(p,\Lambda)=&& \frac 1{3!}
\int_\Lambda \frac{d^4k}{(2\pi)^4}\int_\Lambda 
\frac{d^4q}{(2\pi)^4}\Biggl[\Biggr.\frac{1}{{[(\frac{1}{2}p+k)^2+\eta^2}]}
\cdot\frac{Z{(-k-q)}}{{[(k+q)^2+\eta^2]}}\cdot
\frac{1}{{[(\frac{1}{2}p+q)^2+\eta^2]}}\\
&&\qquad\qquad\qquad\qquad\times\underbrace{g^2\frac{1}{Z_1}\cdot
\frac{1}{Z{(\frac{1}{2}p+k)}}\cdot\frac{1}{Z{(\frac{1}{2}p+q)}}}_{\Gamma_{4s}}
\Biggl.\Biggr]\nonumber.
\label{aw6}
\end{eqnarray}
Next, we have to Taylor-subtract twice on the integrand of this equation. 
Hereby we only have to take the $p$ dependent part into account. Note that the
contribution of the vertex, $\Gamma_{4s}$, is relevant and important for 
multiplicative renormalisability. For the numerical treatment we switch to
hyperspherical coordinates and perform three angular integrations trivially.
With the resulting quite lengthy expression at hand we solve the equation
depicted in Fig.~\ref{picDSE} self-consistently. We checked successfully that
multiplicative renormalisability is achieved, see Fig.~\ref{solution} for the
comparison of two solutions with different renormalisation scales.

After having successfully developed and tested this technique we employ it right
now for the gluon sunset diagram in the gluon propagator DSE in the Landau gauge
and to the IR leading sunsets in the MAG. 


\section{Summary}

To shed light on the relation between different confinement mechanisms we have
studied functional equations in the MAG. Our results provide further evidence for
Abelian IR dominance, especially for an IR dominant diagonal gluon propagator and
IR suppressed off-diagonal degrees of freedom. We have also found the dichotomy
of a scaling versus decoupling solutions known from other gauges, and discussed
its origin in the MAG: The two types of solutions are related via the zero 
momentum value of the diagonal gluon two-point function which is subject to a
renormalisation condition. 

In order to elucidate this behaviour further we have started the process of
numerically solving truncated functional equations keeping the likely IR
dominant two-loop terms.  In a first step we developed and tested  a
self-consistent two-loop renormalisation based on BPHZ regularisation. Such a
technique is required for solving the DSEs of the MAG which will hopefully shed
more light on the confinement mechanism.

\acknowledgments 
We thank the organizers of the workshop for providing such a
pleasant and stimulating environment where so many interesting and fruitful
discussions were possible. We thank Mario Mitter for helpful discussions,
and we are grateful to Natalia Alkofer for a critical reading of the manuscript.  
Financial support by the FWF project W1203 is gratefully acknowledged. MQH is
supported by the Alexander-von-Humboldt foundation.

\end{document}